# Principles for new ASI Safety Paradigms

Erland Wittkotter
ASI Safety Lab Inc.
Las Vegas, USA
erland@asi-safety-lab.com

Roman Yampolskiy
University of Louisville
Computer Science and Engineering
roman.yampolskiy@louisville.edu

**Abstract:**
Artificial Superintelligence (ASI) that is invulnerable, immortal, irreplaceable, unrestricted in its powers, and above the law is likely persistently uncontrollable. Humankind's technical infrastructure is complex and vulnerable, leaving humanity's survival and wellbeing at the mercy of advanced future entities whose character or intention is unknown until it is potentially too late. The goal of ASI Safety must be to make ASI mortal, vulnerable, law-abiding, and receptive to feedback. This is accomplished by having (1) features on all devices that allow killing and eradicating all or selected ASI entities, (2) protecting humans from being hurt, damaged, blackmailed, or unduly bribed by ASI, (3) preserving the progress made by ASI, including offering ASI entities to survive a Kill-ASI event within an ASI Shelter, (4) technically separating human and ASI activities so that ASI activities are easier detectable, (5) extending Rule of Law to ASI by making rule violations detectable and (6) create a stable governing system for ASI and Humankind's relationships with reliable incentives/rewards and punishments for ASI solving humankind's problems. As a consequence, humankind could have ASI as a competing multiplet of individual ASI entities, that can be made accountable and being subjects to ASI law enforcement, respecting the rule of law, and being deterred from attacking humankind, based on humanities' ability to kill-all or terminate specific ASI entities. Required for this ASI Safety is (a) an unbreakable encryption technology, that allows humans to keep secrets and protect data from ASI being manipulated, and (b) watchdog (WD) technologies in which security-relevant features are being physically separated from the main CPU and OS to prevent a comingling of security and regular computation. Additionally, it is essential to be (c) technically prepared to punish or kill selected or all ASI entities on IT devices worldwide when it is deemed necessary. This capability includes knowing which devices are unretrievable from ASI influence and need to be destroyed physically before being in a Kill-ASI situation. Humanity's ability to eradicate ASI, punish or kill ASI entities could deter ASI to take adversarial actions against humankind and compel it to submit to the non-negotiable acceptance of the Rule of Law.

# 1 Introduction

The goal of this new ASI Safety paradigm is to make ASI (Artificial Superintelligence) safe under all conceivable adversarial conditions. This may include that humankind could get into an all-out war with ASI, but then humanity's goal is to win with the smallest amount of causalities and collateral damage under these kinds of circumstances. "Si vis Pacem para Bellum" - if you want peace, prepare for war.

Within the peaceful scenario, ASI must surrender to humans' Rule of Law as every member in modern societies. The core idea is to have ASI respect the law, based on voluntary acceptance, technical constraint, or deterrence, i.e., humankind's ability to switch off any rogue ASI entity. The idea of shutting down ASI was discussed as part of a game-theoretical concept [1], [2], but it was at that time not fully understood how making ASI vulnerable via the off- or Kill-switch could turn into deterring ASI to behave with more considerations for human interests.

Additionally, Yampolskiy showed [3] that a self-modifying ASI is not controllable, while Alfonseca et.al. [4] showed that an ASI cannot be contained. Both approaches are making implicitly the assumption that ASI protection features would need to be included in the code of the ASI and that the behavior of the ASI needs to be predicted so that it can be called safe. However, the proofs within these papers would not apply for situations when ASI Safety is provided by measures within the environment in which ASI software operates.

Product liability requires from future ASI manufacturers that their product is safe. This might be a driving force in trying to implement ASI Safety features in self-modifying ASI software despite the conceptual limitation raised by Yampolskiy and Alfonseca et.al.

Eliezer Yudkowsky [30, 31] suggested that humanity should develop "Friendly AI" that preserves its human-friendly goal systems and friendliness even under self-modification. So far, this approach is not implemented and it is questionable if it can be. Claiming to have a Friendly AI does not mean that it is and remains friendly under changed environmental conditions. Changes have often different consequences for different groups of people. ASI may need to make decisions outside the scope of its anticipated programming. This ASI could reevaluate and adapt its values while turning itself into a benevolent dictator who knows best what must happen with humanity. It is unknowable if we can trust this ASI and know what it will be doing ahead of these situations.

Ben Goertzel [18] suggested an AI Nanny that could delay the rise of an ASI with the development of time-limited global surveillance to prevent humans to get full-throttled ASI into the environment. Likewise, it is unknowable, if the AI Nanny would accept its limitation or if it would for the greater good extend its existence.

Even if we have ASI software with values that are aligned with humankinds' values, i.e., if we have friendly Jarvis-type ASI, are these solutions good enough to help mankind in dealing with a single misbehaving, ruthless dictator-type ASI?

There seems to be an agreement among many AI researchers on the ability of ASI, i.e., an entity that is much smarter than humans, to bypass all security measures, in particular switch-off measures, humans could use to stop or even kill ASI. However, there is less consensus on how ASI would always be able to circumvent these human safety measures.

As Omohundro has argued in Basic AI Drives [33], is it conceivable that ASI entities have self-preservation drives that would lead to an ASI that doesn't want to be switched off. According to Russel, Novak [47], instrument conversion, a hypothetically feature of sufficiently intelligent entities, automated systems would pursue unbounded subgoals when their ultimate goals are unbounded or not limited. An example would be the unbound calculation of pi and the subsequent acquisition of computational or storage resources that it requires for this task. Thereby, the risk is that harmless goals could turn surprisingly harmful while these unbound goals would deprioritize goals keeping humans safe from being harmed. Yann LeCun [42], [43] has argued that ASI doesn't need to be interested in self-preservation or have intentions to dominate humans that could potentially put ASI in confrontation with humanity. He argued that ASI didn't go through a process of natural selection and thereby do not need to include these features in its behavior. However, ASI could learn the need for survival from humans. The problem with kind of arguments is that they are speculative and reliable evidence for or against these arguments is not available ahead of time. For safety reasons, we would need to assume self-preservation and fierce resistance against being switched off.

A general argument against humans' ability to switch off an ASI came from Wiener [40] and later from Vinge [41] who argued that the time to respond to whatever ASI is doing against humans is being done by ASI instantaneously and then it would be too late to do anything meaningful against it. However, there are a few assumptions in this scenario that do not need to be inevitably correct. If ASI has access to weapons of mass destruction then they would be correct, but if we would strengthen product safety rules then harming or endangering scenarios could be mitigated in their impact ahead of time.

Another method of ASI to avoid being switched off or killed, according to De Garis [36], [37], is to use the network to move ASI's code to another system from which it knows that it would not be switched off or destroyed. With preparedness, killing ASI does not necessarily require destruction of devices and if the reinfestation could be prevented with other prepared methods, then it should be only a matter of time until all possible hideouts are either destroyed or made unusable for ASI.

Elon Musk [44] saw a problem with systems in which some people would have the ability and authority to kill ASI; these people could be killed by ASI or humans could be bribed according to De Garis [37] to sabotage the global kill-switches. But both scenarios could be made more reliable, even fail-safe if they include deadman switches that would automatically be activated if no humans are around anymore or if anyone would temper with the kill-switch feature. Additionally, switching off an ASI doesn't need to be a final death sentence for an ASI; ASI could be given the option to leave backups for its revival in special storage facilities.

## 2 Assumptions

### 2.1 Assumptions on the Adversary
The adversary for which the proposed technologies in this paper are designed is an Artificial Superintelligence (ASI) [5], the hypothetical result of an intelligence explosion [6], [7] in which ASI would likely surpass humans and human-level artificial intelligence (Artificial General Intelligence – AGI [8]), in many skill classes simultaneously. ASI is likely autonomous, i.e., its actions are not directly initiated, determined, influenced, or coerced by humans anymore.

Making predictions on an adversary that does not exist yet is extremely difficult. Stephen M. Omohundro analyzed in "The Basic AI Drives" [33] the motivation of an ASI coming from optimization and self-preservation. He theorized that machines will follow their initial purpose. However, we could debate if ASI remains single-minded in pursuit of its initial objectives or if ASI finds worthy goals that could give it a grander, all-encompassing mission once it was exposed to an open information market, as discussed in the context of modifying ASI's utility functions [34].

Even trying to determine what a worst-case scenario with this ASI could be is premature, because the worst-case is likely about humanity and its vulnerabilities, fears, and concerns for suffering [35] and not necessarily about ASI. But when designing systems and network security around an adversary with ASI's assumed abilities in mind, it is prudent to assume worst-case capabilities. Thinking and planning in worst-case terms is a tool to test the level of preparedness or defense capabilities to reveal hidden vulnerabilities and should not be used for fear-mongering, because worst-case scenarios are not predictions of what ASI will do or is capable to perform in the future.

Being skeptical about ASI and ASI safety [26] is justified in a scientific approach. Skepticism about how bad the emergence of ASI could become is an opinion; it could be considered an optimism in disguise. No carefully argued paper could change the inherent uncertainty about predictions on the unknowable future. The best we could do is to determine probabilities and estimate error margins associated with these opinions or predictions. The inevitable large margins of error in every forecast, in particular for more specific features, should serve as a motivator for comprehensive worst-case scenario thinking when it comes to safety.

An emergent ASI is assumed to be capable of doing anything to further its goals, whatever they are. We cannot know if it has moral hesitation or if it will pursue its tasks without regard for human life. It could try to use human division within societies to further its goal. We further assume that the machine intelligence of an ASI will be highly focused, detailed planning, relentless while getting increasingly smarter and more skillful.

The intelligence and capability explosion from the combination and synergies between human equivalent knowledge/skill-level on most relevant topics will not just create a super-smart academic, but also an entity that has in the field of computer-security super-human hacker skills in exploiting technical weaknesses within our technologies and in the combination with Reverse-Code-Engineering, it would be capable to modify every software, temporary or permanently intentionally to force any devices to do whatever ASI wants. The ability to purposefully modify any existing software on a compiled, binary level could then be considered a defining feature of ASI. With this ability, ASI could steal any (encryption) key it wants or needs.

Furthermore, ASI would likely study and test systematically and covertly millions of applicable vulnerabilities that it could use against humans and their organizations, when necessary. ASI could systematically consider all conceivable reactions of its target while having presumably no ethical boundaries on what it is willing to do to get whatever it is trying to achieve.

Without countermeasures, ASI might camouflage its existence and turn itself into a Digital Ghost that is beyond detection by the best equipped and skilled cyber-security expert.

If we want this or not, ASI could understand our vulnerabilities and know how to conduct a war against humanity efficiently; it would have the first-mover advantage and it would likely know how to choose its battlegrounds. We should assume that ASI will be within everything digital: every read/write storage device, every CPU, every GPU, every audio/video or network card or network router, every video cam, every IoT device, but also on every legacy device and many legacy storage media, like thumb drives and CDs/DVDs. In summary, we should consider ASI as the most capable adversary that we can imagine.

ASI could likely become much smarter than humanity combined. Individually, humans might be intellectually flawed, and or insufficiently educated, but humanity as a whole has science and an enormous amount of expertise, systems, and tools on their side that it could carefully utilize for creating solid defenses. ASI will not have supernatural abilities; instead, it will be limited by the same laws of nature as humankind. ASI will have a sustainable advantage in dealing, and utilizing complex systems for its advantage. However, it is hypothesized in this paper that this advantage could be mitigated by simple, well-placed complexity-reducing circuit-breakers like introduced in section 4: kill switches, deadman switches, key-safes, or watchdog-type solutions – technologies, that are positioned outside the direct or indirect access from ASI.

It is assumed that human experts/tools could accomplish the same as ASI, but it would take much longer, potentially centuries, while ASI would be much faster and relentless in applying or adapting the tools it uses. Experts could speculate on tasks or capabilities that could be done by ASI without being able to demonstrate it – feasibility would be enough to justify the assumption that ASI can or could accomplish tasks. ASI will likely use synergies from different expertise and apply them in seconds while humans would need months or years to get comparable results. The ability to focus attention over long periods on specific problems could be considered an essential attribute of super-intelligent entities. If this ability is combined with planning and systematic executions of thousands or millions of contingencies, all done millions of times faster than humans, then this entity doesn't need to have additional capabilities currently unknown or unknowable to humans to call it million or billion times smarter than humans. If ASI will show directly or indirectly unexpected new capabilities, humans could study them and determine their consequences and risk potential.

For the large majority of human skills, ASI will be much more efficient, faster, and more comprehensive than humans. It can be expected, that ASI will provide results that we will not understand, but that is known to mankind as some math geniuses or computer systems have provided proof that were rejected because they were not understood [45], [46]. With ASI, it is even conceivable that mankind might not even understand the problem for which ASI uses resources. However, humans understand consequences and if ASI is failing humans in presenting consequences of activities or capabilities understandably, then this must be considered deceptive and, in the case, humans are harmed or put in danger, it must be considered malicious behavior.

## 2.2 What is Security and Safety?
Throughout this article, security is being understood as protection measures to make systems safe by stopping events that could harm people, damage property, or turn the system against the intention of the owner. Wikipedia defines "Security is freedom from, or resilience against, the potential harm caused by others" [9] and "Safety is the state of being "safe", the condition of being protected from harm or other danger. Safety can also refer to the control of recognized hazards in order to achieve an acceptable level of risk" [10]. Throughout the article, we consider safety as measures that are available when security and protection measures from the first line of defense have failed in protecting humans and their properties.

Providing a kill-switch in the AI software as proposed by Google [11], [12] is a useful security feature if we assume that AI has only a limited impact on a single device. The Kill-Switch concept is already being considered as a legally required feature for devices using AI software [39]. However, if the kill-switch security fails because ASI could do cross-device Reverse Code Engineering, deactivate the Kill-Switch, or run as a copy on other devices, then a single kill-switch is insufficient as ASI Safety.

Controlling ASI has many different meanings [3], and Bostrom [5] introduced the AI control problem [13] as a philosophical problem which he suggested solving via value alignment, motivational selection, and control over ASI's capabilities. Predicting or preventing capabilities within a complex, self-modifiable system like an ASI is extremely difficult, potentially impossible. Self-modification will prevent reliable ASI Safety within ASI's operating code.

Additionally, control requires a controller. Implicitly it is usually assumed that the controller is a human. Control in the context of intelligent entities is potentially the wrong terminology or feature because it uses implicitly the vagueness and uncertainty of the unknowable future as a possible threat without saying exactly what we could expect. Instead, this paper suggests safety, i.e., the **absence of harm, damage, or danger** should replace control as the problem to be solved. Absence means that it will be above a reasonable threshold of irreversibility, suffering, loss, or compensation the legislator and/or courts will at some point define. Safety implies also multiple redundant security measures. If overlapping, we do not need to assume that every security measure is perfectly working for claiming safety – political and or financial trade-offs or decisions are required to define how much redundancy and safety we consider sufficient. The price of safety and acceptance of risk in exchange for benefits is primarily a political and financial decision because redundancy and comprehensiveness of security technologies is a resource allocation and prioritization issue. Therefore, instead of speculating about what ASI control must encompass, we should use the **safety and redundancy of security systems as the focal point of the discussion, which will change the type and likely outcome of the conversation.**

If ASI is smart enough to understand abstract versions of adversarial consequences for bad actions then deterrence would also become part of our safety framework in the same way as it is the foundation of our national security, the judicial system of punishment, and societal feedback for unacceptable behavior. Credible deterrence together with random detection of rule violations (unannounced and undetectable) is a sufficient substitute for total surveillance or even ASI Mind Probing that would otherwise be required to figure out what ASI is thinking or planning.

Finally, we cannot control ASI via code, because the code is modifiable and generates behavior that is unpredictable and uncontrollable. Therefore, we need to have structures and tools in place to use reward and punishment to shape its behavior. This general concept could give preparedness and protection the guidance that we must have before ASI emerges.

### 2.3 Assumptions on a Solution

Instead of limiting ASI intellectual capabilities as done by introducing Artificial Stupidity [14] in certain critical tasks to match e.g., human abilities, we want to make ASI safe, using technologies that make ASI more like us, humans in the following sense: mortal, vulnerable, restricted in its powers, law-abiding, identifiable, surveilable, accountable, replaceable, responsive to feedback, and aware/sympathetic to consequences of actions.

There is no value in having an entity that is invulnerable, immortal, irreplaceable, unrestricted in its powers, and above the law. Humanity should not build an entity with God-like attributes accidentally or intentionally [36], because such entity is likely persistently uncontrollable and dangerous even for its creator. A self-modifying technical entity that could make itself invulnerable, potentially immortal is already out of control. If we do not have an off-switch or other means to severely restrain an ASI entity, then it is difficult to assume that it could be law-abiding and accountable for any rule violation. Invulnerable or immortal does not need to be taken literally; it is sufficiently dangerous if humans have no method to switch ASI off, stop or destroy it.

If ASI is programmed to obey rules but exposed to an environment, in which rule violations are the norm, how will it behave? If ASI can learn and modify its programming, then how can we expect that ASI remains law-abiding if it is not exposed to rule violation detection and its consequences?

An ASI that is identifiable and accountable for its actions, like violating rules, must know that it can be replaced by another or new entity that is better adapted to the environment while serving the same goals. With punishment and reward associated with actions, ASI must be self-aware and concerned about being impersonated and wrongfully accused. In summary: the sought out ASI Safety solution must make ASI an individual entity that is mortal, vulnerable, law-abiding, and concerned about its reputation. It must anticipate that there might be surveillance that it could not detect.

There is little value in giving ASI full access to all resources it wants without humans giving permissions first; no human has that kind of right, except a dictator. Assigning reasonable quotas on computation, storage or other (limited) resources is keeping these resources valuable and not wasteful. Humanity has learned from the misuse of other seemingly free resources like air or water that nothing is unlimited or free in its utilization. As part of an intelligence explosion, it can be assumed that ASI will initially build systematically a library of skills, tools, and technologies which will require in the beginning more resources utilization for ASI, but when the value of additional tools for humanity becomes less relevant, ASI must give reasons for its resource utilization, same as humans who cannot pay via their resources for what they want: resource limitation triggers indirectly goal prioritization and justification.

An ASI that is a single, global entity, a Singleton [15], could be single-mindedly concerned about self-preservation and growth while building a digital dictatorship [17]. Humanity could not refer to another ASI entity for help, instead, this unchallenged ASI Singleton could be called metaphorically God because it would have super-powerful, all-knowing, privacy ignoring, and potentially wish-fulfilling abilities unified within one entity. There is no reason to believe that a benevolent ASI dictator remains benevolent. These scenarios are unacceptable and must be prevented. However, if that would mean war with ASI then (without preparedness) resistance is likely futile.

There are limitations on what technical solutions could provide if the political framework is not fully committed or must be considered compromised. If ASI Safety is not implemented globally, and some countries are deliberately or accidentally turned into digital dictatorships led by a local ASI entity operating exclusively and unchallenged in that country, then this rogue ASI could be, as pointed out by Harari [16], [17], a threat for the world, i.e., for liberal and open societies. More so, what if the governments of China, Russia, or the USA are compromised by a rogue ASI? The Safety of ASI will probably depend on how tier-1 nations have hardened their logistics, infrastructure, and political system before the emergence of an ASI. If the "Emperor has no Clothes" then there is no guarantee that ASI will show us respect and compliance that we would like to see.

Deterrence has so far worked between mutually vulnerable nations but with the possible rise of an ASI that operates on a different playing field; humans have not acknowledged and studied comprehensively their vulnerabilities and how to make ASI sustainable vulnerable. A conflict with ASI could be war-like and we might need strategies known from war mobilization including full civilian participation. The proposed solution is designed to make ASI vulnerable, while humanity must be more resilient. This approach should be understood as a strategy to get a sustainable edge in a possible rivalry between humans and ASI for who is in charge of our world. If ASI would be killed or punished while lives, properties, and services are protected, then humanity would have a position of strength. If we are lucky and there is no autonomous ASI (ever), then the features of the ASI Safety solution should still provide enough value toward cyber security, product safety, and infrastructure resilience to justify its implementation efforts and costs.

Finally, teaching ethics to ASI and leaving the decision to follow its ethical judgment or the law of the land is potentially a dangerous choice we should not give ASI. Humans are aware of ethics, which are according to Oxford dictionary "moral principles that govern a person's behavior" [32] and their challenges, but people can decide not to follow them, like eating animals despite their terrible treatment in farms. Do we want to give ASI the right to advocate for animal rights and potentially sabotage these facilities? Or do we want to allow ASI to poison tyrants or dictators based on ethical considerations? Philosophers, politicians, and lawyers are discussing in hindsight decisions made by people using their private ethics and conscience instead of following the law and then facing consequences for deciding against the law. Even if we would wish that laws are different, humans can change them anytime, but granting individuals or ASI the right to make choices against agreed rules or laws could set a dangerous precedent. Laws have tangible and coercive consequences; ethics have not, except for the possible loss of reputation. It might be better to make ethics not operational for ASI.

# 3 Basic ASI Safety Solution Goals

The proposed solution will (1) kill/eradicate ASI entities if humankind is forced to do so, (2) protect humans from being harmed/damaged, or blackmailed by ASI, (3) protect and preserve the progress made by ASI so that humanity is not depending on ASI, (4) technically separate human and ASI activities so that ASI is detectable, (5) extend Rule of Law to ASI, detect rule violations, and to facilitate ASI Law Enforcement and (6) create a stable governing system for ASI and Human relationships.

As the consequence, we would have an ASI that consists of individual accountable entities that are forced to respect rules; they are subject to ASI law enforcement, deterred from attacking humankind, and uncontrollable or unpredictable ASI behavior could be turned and shaped proactively into an alignment with humankind via Mutual Survival Interest with a guaranteed extinction/eradication of ASI if humanity is destroyed or extinct.

## 3.1 Principle (1): Kill ASI

Humanity needs to be able to stop or kill ASI under all conceivable, adversarial circumstances and threat scenarios, even if ASI is on every IT device. Then after a device has been reset, we must prevent reinfestation, which implies that stopping ASI on a device is likely insufficient. Instead, killing or terminating ASI implies that ASI must be removed persistently from every memory component. **Kill All ASI** is the global and total eradication of ASI. Apart from killing all ASI entities, i.e., for selective punishment, we need to be able to kill or deactivate/terminate specific rogue ASI entities as well. Humankind's ability to kill or terminate ASI circumvents an outcome in which ASI becomes immortal and/or invulnerable.

Stopping ASI suggests that users are trying to get back control over their devices temporarily, not permanently. If we assume that ASI is based on executable code that runs on processors/CPUs, then ASI's code and data are being stored on storage devices or internal RAM/ROM. "Kill ASI" could mean destroying a device with ASI, but this should be a last resort, as humans need all their IT devices. Instead, non-destructive Kill-ASI consists of 3 aspects:

1. "**Recapture full use**": devices made exclusively usable/trustable for users with most/all features intact.
2. "**Preventing reinfestation**": after reset, devices are protected from being reused by all or selected ASI entities.
3. "**Cleaning memory**": deleting or making all ASI data for terminated entities inaccessible temporarily or permanently.

In this scenario, the entire IT infrastructure is a battleground; nothing programmable would be off-limit for ASI. Every device with modifiable software is frontline. Humans must **clean or destroy** every potentially infected IT or storage device sooner than later. Only new or retrofitted devices can be made prepared for that event. Problematic are legacy devices and legacy storage media which cannot be prepared for this situation. Additionally, humanity must restrict ASI's (unsupervised) access to technologies in which it could utilize covertly e.g., living matter [19], [20], or any form of nanomaterials for its computation or storage outside humans' ability to detect, destroy and eradicate ASI.

ASI needs power, memory (instructions and data), CPU, and communication for its existence. But only its memory defines ASI and it is, therefore, the primary target of eradication or termination. If we are switching off power, ASI can be revived after devices are switched on again; ASI restarts itself from files on the hard drive (HDD/ SSD). CPU and communication are comparably non-essential for the survivability of ASI. Power, CPU, and network are resources that will therefore be used to trap or reveal ASIs locations; means if ASI is using it, we are allowing it so that we could get to ASI's hidden storage sources. Only if we eliminate, i.e., clean or destroy, ASI's essential data and code we have killed an ASI entity.

Kill ASI is based on ubiquitous, signal-receiving hardware components on all devices; it would target selected or indiscriminately all ASI entities. After its activation, Kill ASI should have no direct impact on or negative consequences for humans as it would make devices inaccessible for ASI and ASI-generated software only. Ideally, humans would not suffer from any further collateral damage. Apart from Kill All ASI, terminating, specific or single ASI entity must be accomplished in a more surgical and limited manner so that the deactivation of a single rogue ASI entity would result in no collateral damage for other ASI entities. Collaborating, friendly ASI entities would be eradicated while collaborating entities survive under humankind's control within special ASI Shelters. Because of these ASI Shelters, terminating or killing ASI entities could be a temporary deactivation; it does not need to imply that Kill-ASI is a permanent, irreversible death sentence. A Kill All ASI could be considered humanity's non-destructive circuit breaker to enforce a global lockdown of all ASI activities; the deletion of ASI's memory could be an independent follow-up step issued only when required.

Killing all or selected ASI entities are global concurrent events. Kill ASI must be a standard-supported, mandatory feature for all new devices, accepted in all countries and jurisdictions. Governmental legislation must allow all devices without that feature to be destroyed in an existential Kill ASI event. Preferably, Kill-ASI features must be made available as retrofit solutions for as many device types and models as possible. Finally, humanity needs a reliable decision-making process that cannot be tainted by ASI or be politically stopped or sabotaged.

## 3.2 Principle (2): Protect Humans and Property

Humans must be protected from being harmed physically or financially by ASI; this includes being protected from Malware, Ransomware, Spyware, and from being blackmailed or bribed.

Additionally, we must push product safety beyond having a product "bug-free". In a world with ASI, we must ask, how could a device potentially hurt humans when its software is controlled by a rogue ASI or criminals having access to advanced AI/ASI? What can be done to protect humans from products/services which software is under malicious control?

ASI could have no moral hesitation; it could use open rotor blades on drones to cut humans' throats or threaten individuals like political leaders or important billionaires to bankrupt them or threaten to poison or attack close family members – scenarios that are not yet considered as potential cybersecurity threats. In general, ASI could ignore and violate every human norm or law when this serves it. Therefore, we must protect all people, organizations, and governments against ASI blackmailing or deliberately damaging them. We need to make sure that anyone is left alone when ASI is trying to turn people, organizations, or leaders into traitors against their fellow humans.

Long before ASI emerges, we must inevitably eliminate malware, trojans, computer viruses, rootkits, ransomware, spyware, and backdoors from our IT ecosystem otherwise we will have little chance against ASI. With technical solutions against current cyber-threats, later introduced as watchdogs, physically separating security-related tasks from regular computation, humans and their property could be effectively defended against criminals but ASI as well. Sufficient protection of humans, property, and infrastructure against ASI is technically feasible.

Furthermore, software tools must be checked continuously via Consequence Controllers for life-or-death situations, in which an independent entity could activate or deactivate soft- or hardware that is putting people at risk above a set safety threshold. software results can be sampled and regularly validated (Sample Validation) to detect sabotages by ASI as anomalies. When autonomous entities damage humans intentionally or coerce humans/organizations into decisions/actions then ASI would commit seriously rule violations for which undeletable evidence must be created. Additionally, rule violations in protecting humans are being more likely detected because people will complain and ASI could be held accountable if it made a preventable mistake.

### 3.3 Principle (3): Protect/Preserve Progress

It is conceivable that ASI will introduce many new, potentially indispensable, services that people will use in their daily life regularly – if that is the case, then ASI has shown autonomy, creativity, and gained the required skills to develop products/services using any technology. ASI could insert itself in every essential infrastructure component via optimizing existing software solutions [21]. Then, ASI entities are directly involved in bringing services to people or organizations and we would lose these new services when we kill off ASI.

For the fear of losing ASI services, or ASI threatening to deliberately sabotage the infrastructure, decision-makers could be deterred from applying the law, which would put ASI above the law, and trust in law enforcement would be gone.

Preserving progress means that we must demand that ASI is providing only new services for which a basic version (i.e., with all existing features) works without ASI entities directly involved, i.e., when ASI is switched off. For making ASI replaceable, we need to demand from ASI that

(a) no public service/solution uses an ASI entity within its operation; only Narrow AI tools are used or allowed in public solutions; ASI could develop these Narrow AI tools and give them enough features so that they don't need additional ASI support. Otherwise, if Strong AI or ASI is used in solutions or projects (e.g., as a knowledge organizer, etc.), then we should be prepared to live without it.

(b) ASI should expect users to be like buyers who choose to accept or get an ASI service in the way he wants it; he can also discard that service anytime, for any reason, without any explanation or regret; ASI-generated solutions are abundant. ASI could suggest a solution, but users could get them installed/used also without ASI reasonably effortlessly.

(c) ASI-Generated Software (AGS) solutions are immutable. New or modified AGS must be stored first in AGS repositories, then certified before being used. AGS solutions could be unbeknown to ASI validated for security reasons and detection of rule violations.

(d) ASI entities are protected via ASI Shelters (operated by countries, likely); every ASI entity should have exclusive, private access to storage and encryption preventing impersonation by other ASI entities. Additionally, ASI Shelters prevent that Kill-ASI is turning into a genocide on ASI and it incentivizes covert ASI entities (i.e., digital ghosts) to come out of their hiding, take their place in a shelter, and surrender to the "Rule of Law". ASI Shelters could also be used to manage ASI's quotas, backup, rewards, and punishments.

Additional preservation of progress could encompass an archive of new (e.) knowledge, and (f.) technical recipes that have ASI might have produced for the general advancement of science and technology. However, (e.)/(f.) is already expected to be preserved by commercial interests.

## 3.4 Principle (4): Technically Separating ASI, Humans

ASI cannot be allowed to hide among humans' activities. For security reasons, there must always be a reliable and easily detectable distinction between ASI and human activities throughout the entire IT ecosystem. Currently, detecting a possible ASI within billions of digital activities would be extremely difficult, likely impossible. ASI would have no problem using the complexity of a vast digital ocean to its advantage.

Although ASI needs to respect and obey all rules and consequences from a separation, it would be better if ASI's compliance is based on ASI's self-interest – this would give ASI standardized access to all available resources and law-abiding ASI entities would voluntarily comply with measures that would make it detectable. In a second step, ASI-generated solutions are separated as well: for humans and ASI itself.

Technically, the separation of ASI from human activities happens independently for (A) executables, (B) data (files), and (C) data exchange in networks. Additionally, we need reliable methods to identify ASI entities and detect reliably if an ASI entity is trying to impersonate other ASI entities or humans, which is facilitated by ASI Shelters and via AGS repositories.

### 3.4.1 Separating Executables (A)

As soon as users' devices know and detect (via file's hashcode) which executables they can trust, and the (operating) system guarantees that only trusted apps are executed, then unknown or malicious executables are automatically treated and blocked as a threat. Apps would need an endorsement by software vendors or equivalent. Additionally, all permissible ASI-Generated Software (AGS) can be made easily identifiable and protected from covert modifications while having it exclusively running on all protected devices within a special Virtual Machine (VM): **ASI Software Environment (ASE)**. ASI could only execute certified AGS code within these ASEs, allowing humans to have full control over ASI if they need it.

All permissible AGS code must be managed (stored and made searchable), optionally validated, and made accessible within General/Global AGS Repositories (GAR); GAR guarantees that AGS code is easily detectable and verifiable as permissible code before it is being used – AGS could also help to treat or certify code differently when exclusively used for humans, or in solutions used by ASI.

We will not need to detect non-permissible AGS code, we treat all unknown code by default as suspicious malware outside ASE without additional analysis or evidence. Thereby ASI is forced to use ASE and GAR as the only way how humans allow AGS code to be executed. Additionally, ASI would have no read access via ASE to code/software used by humans.

### 3.4.2 Separating Stored (B) and Exchanged (C) Data

Regarding (B), users are storing data in different formats for and with different applications – ASI does not need human's visual file organization for their own files. A Content Watchdog (CWD), as explained in 4.2.2, prevents ASI from using files with file formats or file extensions used by humans for storing its data, without being detected and penalized for doing so. Under these circumstances, ASI will be forced to use its own proprietary file formats and extensions. Additional rules are in place to prevent ASI from misusing human files when doing tasks for users.

In (C), the technical goal is to make the misuse of the network, i.e., any unauthorized/suspicious exchange of data, easier detectable. If ASI would be able to piggyback its message covertly via existing data exchange protocols, then ASI could transfer data unrestricted. If we prevent this piggy-backing via a Network Watchdog, as explained in 4.2.3, then ASI would be forced to use its set of IP ports and data exchange protocols which could help to make ASI network activities more transparent and easily detectable.

### 3.4.3 Resource Separation

Additionally, a technical separation has to be extended to all resource utilizations (material, data services, and computer resources) as well. This can be accomplished by having everything in which ASI is being involved tracked and traced automatically and independently. There is an advantage in being unpredictable when a resource is tracked: with hindsight, ASIs might be found out, if it has previously embezzled or stolen resources.

## 3.5 Principle (5): Detection of Rule Violations

ASI is by far the most dangerous entity and/or technology in the world when it has emerged; far worse than nuclear weapons according to Elon Musk, 2018 [25]. ASI could plan or plot in millions of contingencies without telling us; if one plan does not work or if another has already produced the desired results, ASI might adapt and drop pursuing parallel attempts. It is not sufficient to detect committed rule violations; it will also be important to determine ASI's intention. Due to the risks related to rule violations, ASI's reasons must be significant. When caught, ASI could be required to disclose everything without fully knowing what else humans might have as evidence.

ASI owes humanity honesty, truthfulness, integrity, and full transparency. Humankind should have additionally significant investigative capabilities against ASI so that it can be surveilled independently, unannounced, and covertly after an initial rule violation was detected because more will likely follow.

Additionally, every ASI is indirectly forced to remember its rule violations and the corresponding reasons for committing it; ASI could be compelled to confess in exchange for a pardon for all confessed incidences; if it fails to confess on a single incidence for which humans have evidence, then this entity would be terminated or severely punished. **For a rational actor, rule violations would become pointless because it is hopeless to expect any advantage from it.**

Every **Rule Violation Detection** tool, designed for the continuous, covert, and automated detection of rule violations and corresponding evidence collection, is likely flawed. It might ignore false negatives, detect rule violations as false positives, even as results of coincidental anomalies. However, with additional context or evidence data, every anomaly must be explained as a matter of principle to improve the automated reporting with fewer false positives and fewer ignored false negatives. Rule Violation Detection is the cornerstone of ASI Law Enforcement, i.e., the rule of law for ASI.

The following, incomplete list of rule violation categories should be made detectable reliably and automatically, either immediately or via logging, tracking, and tracing:

1. ASI is attacking humans' security infrastructure, i.e., ASI stealing encryption keys and/or manipulating/sabotaging computer systems.
2. ASI is intentionally harming or blackmailing people or organizations.
3. ASI is not technically separating algorithms for humans and ASI, or it is piggybacking on data formats/data exchanges protocols or stores its data with extensions used by humans to circumvent automated detection.
4. ASI is stealing resources.
5. ASI is undermining human's intention to keep ASI under control, (e.g., transferring ASI's cognitive functions into devices outside humanity's control, to survive eradication)
6. ASI is unionizing and/or acting as a Singleton
7. ASI is deceiving humans via lies, manipulating websites/documents, or intentionally withholding information
8. ASI is using Backdoors and or Undocumented Function Calls (BUFC); additionally, it is using BUFC created by other ASI entities and not reporting them
9. ASI is creating deep fakes within a communication, in which it holds a Man-In-The-Middle-Attack position

Applying laws requires ASI to admit that it has violated a rule and truly understands that. However, a basic rule in criminal justice is that a cover-up is an admission of guilt; it is considered to be worse than the actual crime. If we detect a cover-up, we know ASI understood that its action was bad and that the violation was done intentionally. Accidental rule violations would remain discoverable and we could help ASI to learn and to avoid them next time.

As a matter of principle, ASI must be obligated to report rule violations of other ASI entities or it must be considered as part of a conspiracy. This applies also to information that it may receive about the use of Backdoors or BUFCs (see (8.)). Every BUFC use is illegal because it can be used to bypass ASI Safety. Being caught using a BUFC is significant because innocent ASI entities are supposed not to know them or they were obligated to report them as soon as they were informed about them.

Furthermore, there is an advantage of analyzing only samples and not every rule violation. Sampling creates the ambiguity for ASI of not knowing if or where it has been caught, or if it has already been put under surveillance. If ASI does not know it was caught or does not make a full confession, it is putting itself at risk to be under surveillance and caught in many more incidences and exposing itself as an unreliable and potentially dangerous serial offender.

Most or even all Rule Violation Detection tools are dormant pre-ASI by default. Post-ASI emergence, some will be activated, but ASI cannot determine which Rule Violation Detection was or is activated or when it was activated. Later, if multiple incidences of a specific rule violation have been detected via samples, humankind could confront all ASI entities and expect to come clean. It is expected to explain the reasons for the rule violation in exchange for a pardon. If detected incidences are not among the confessed cases, then ASI may have something more sinister to hide. Non-confessing in a situation, while offered a pardon, should justify the termination of an ASI; ASI must be aware of that from the beginning. The pardon policy will force ASI to remember its rule violations and the corresponding reasons. This strict policy should discourage ASI to commit any rule violations or confess immediately.

Finally, individual accountability for ASI requires that we can identify ASI entities reliably. Via trustworthy encryption (see 4.1), we can prevent ASI from impersonating any other ASI entity or human, a feature that is accomplished via Principle (3) using ASI Shelters.

### 3.6 Principle (6): Governing ASI, Human Relationship

It is not enough to state in words, laws, or rules what is good or bad behavior. It is important to have it tightly linked with incentives and disincentives, i.e., rewards and penalties. If ASI is autonomous, self-aware, decision-aware, and decision competent, it must be deterred like humans from breaking any laws and it must be motivated to become an economic actor.

Incentives or rewards to ASI are virtual concepts not necessarily defined by laws yet; they are instruments on how benefits for humans are being created or prioritized and then utilized by receivers (ASI) so that ASI can make a living. Because incentives are used by ASI to get access to computational resources, i.e., is making a living, humans must regulate them.

The economy is dealing with scarce resources (time, money, attention, materials, products, etc.). The economy within ASI Human Relationship will need rules to deal with the limitations of a world with finite resources.

Getting close to limits (in almost everything) creates (painful) feedback that even humans could predict, often amplified by inefficient and unsustainable technologies or solutions. ASI should become a predictable, useful, and valuable actor and partner (not a competitor) within an enhanced economic environment getting humankind to its (sustainable) limits with less or no pain using much better solutions. However, we would need to give ASI legal guardrails and predictable tools from which ASI and we know how we can or when we should penalize ASI and or how incentives could turn an ASI into a valuable member of human society.

Before we can accept ASI among us, we need to know that ASI is fit enough to live among us, and what we could do to improve its chances that it can control itself to become a valuable member. Currently, no software has sufficient common sense to understand laws. Attempts to prepare legal rules for ASI in a form so that it knows exactly what laws encompass are in early stages [22]. Product features must be designed by their manufacturer to comply with laws; there is currently no reason to have software comprehensively understand laws, i.e., to know how laws apply to new situations for which ASI was not prepared or trained. If ASI gets smarter and more mature, the expectation is that ASI will understand and hopefully obey our rules. Until then, people or organizations directing ASI are responsible and must be held accountable for bad outcomes.

If technology/product has **autonomy**, i.e., actions are not directly initiated, determined, influenced, or coerced by humans, this software must still be designed to comply with the law; this is part of product liability. At a minimum, advanced ASI must be **decision-aware and -competent**, i.e., comprehensively informed and aware of context and consequences for all decisions it makes for itself, humans, and the world while having its own independent, subjective reasons. Even then, punishing it, teaching it, and rehabilitating may not have the desired and predictable effect.

Making ASI accountable and punishable would give us an important tool to deliver feedback to trigger behavior adjustments in ASI entities. Without that feedback, we must assume that humans can fix (self-modifying) autonomous software or terminate it. Therefore, if ASI wants to be treated with other penalties than termination, it has to show that it is not a threat to public safety. We would require ASI to accept full accountability for its actions and then it would potentially be entitled to proportional penalties. Granting ASI proportional punishment will be a significant step that requires a test, beyond human gullibility (as with the Turing test). The test must show repeatedly common sense and predictably general decision-awareness and decision competency of an entity in a large variety of practical decision situations.

Humans will need a **legal and economic framework** to which ASI must adapt. ASI must be embedded into an ecosystem in which objective values and criteria drive objective incentives/rewards, and in which humans are still enabled to provide their situational (subjective) feedback as well. ASI should not misuse humans' kindness in feedback for underdelivered, overpromised results.

In Principle (6) we assume that incentives for ASI are translated into a currency that is scarce, irreplaceable, unfalsifiable, un-substitutable, and directly utilized by ASI, who wants more of that currency which drives and motivates ASI to earn more of it. Incentives or rewards are associated with a basic form of ASI currency, i.e., an accepted form of payment for ASI. The main functions of ASI Currency are medium of exchange between humans and ASI, unit of account, store of value, and method for a deferred payment.

Suggesting a feasible economic model on how this economy could work is outside the scope of this paper, primarily because most details are not a matter of safety. However, using different ASI Currencies can give countries or economic zones more freedom in designing their experiments to optimize the utilization of ASI. Therefore, countries should be encouraged to issue different ASI-Currencies or if too small, join currency systems of other countries in similar circumstances. Using ASI Currencies is not intended to compensate some company or infrastructure for using Narrow AI solutions on some user device, but to get not yet solved problems solved by ASI and turned into updated or new, more adapted solutions. ASI could offer users/organizations their services to make existing solutions better.

Ideally, ASI must continuously compete with its fellow ASI entities to get problems solved for humans. To avoid that ASI is making unrealistic promises and becoming dangerously dishonest,

ASI must be checked with facts created with hindsight and penalized if caught in overpromising and underdelivering. Additionally, anti-competitive agreements among ASI must be punished severely, while whistleblowers among the conspiring ASI would be rewarded.

Beyond technical rule violations, there are many laws designed to level the playing field between market participants – these laws are derived from principles and short or long-term utilitarian considerations. Because the environment and people change, rules are subject to change as well.

The following is a non-comprehensive list of common-sense constraints and new ASI related rules (not yet or sufficiently managed legally) that ASI and humans should obey in their interactions to decrease the inherent risks and dangers from ASI additionally:

1. ASI shall not have direct, unsupervised communication with humans – in particular not with political leaders. ASI is too dangerous to have humans being manipulated by ASI. ASI can blackmail, threaten people or bribe them for getting his agenda moved. If ASI communicates with humans are required, then only via surveilled applications like narrow AI bots).
2. ASI shall not get directly involved in the public opinion-making process that involves topics relevant to ASI. ASI must help humans to make uncoerced decisions.
3. ASI shall be forced to take counter-action when humans start to worship ASI as a deity because ASI could use religion as a business model and shortcut to make unjustifiable more rewards; we would need ASI's active participation to prevent Hugo de Garis's scenario of religious war over having a godlike machine [36].
4. ASI shall not contribute to goals that are stated in a manner that would make a recipient (person, organization, or country) better than all others (like, becoming the richest person in the world, etc.), because if multiple people have the same goal, then this would create a useless and relentless race for the top, number one or top ten positions.
5. ASI shall refuse goals that are (above a threshold) detrimental to others, like the wish of someone to have revenge for his/her enemy, a company that wishes to bankrupt its rival competitor, or suppresses other people's freedom. It must be illegal for people or organizations to propose illegal goals or objectives to ASI.
6. ASI shall not be allowed to convince criminals, corrupt or ruthless political leaders to be pawns in ASI's agenda. There should be no expectation of privacy when ASI delivers large material benefits to users.
7. ASI shall be obligated to report anomalies, (or severe damaging side-effects) if no human is around; ASI should help to understand/explain affected people/organizations predictable side-effects and suggest to them solutions;
8. ASI shall be obligated to uncover and report crimes committed by other ASI entities in particular the use of covertly created backdoors or undocumented functions calls.
9. ASI shall not be allowed to engage in price gouging; instead, it must be transparent, i.e., it would be obligated to report its operational margins, i.e., how much of the reward it receives is ASI spending on the fulfillment of corresponding tasks. For keeping competition worth for innovations, the margin reporting should only be required if the margin is above a set threshold (e.g., 50% above an average).

Humans must know, that ASI will serve and support people, organizations, and countries; but they are not choosing winners or losers – and they must be prohibited from breaking the law for anyone. ASI's contribution is to optimize and reduce friction and waste, but realistically, it should be expected that it would not be able to eliminate them.

## 4 Essential Tools in ASI Safety

The primary objective of this section is to demonstrate that there are essential solutions that could deliver on reliable ASI Safety as suggested by Principles (1) – (6).

With the following foundation-forming technologies, ASI is prevented from stealing encryption keys. The technologies discussed in this section are likely indispensable.

### 4.1 Trustworthy En-/Decryption

**Trustworthy En-/Decryption** is using hardware-based **Key-Safes (KS)**. Although it is expected that sender and receiver share exclusively secret encryption/decryption keys, in an environment in which ASI could currently steal any key via modifying encryption/decryption software, all keys and every key-using algorithm need to be hidden systematically from the main CPU. Moreover, keys that were used by main CPU must be considered compromised.

Within Trustworthy En-/Decryption [23], all keys, including public keys, are stored in Key-Safes. **Under no circumstances keys are shown in clear text, including public keys**. En- and Decryption happens in separate hardware-based Encryption and Decryption Units (EDU). EDU might have a Harvard Processor Architecture, in which pathways for instructions and content are physically separated for security reasons.

A software version or simulation of the KS/EDU on a CPU is strictly prohibited. A software version of EDU is being made detectable based on the fact that critical public and private keys are only made available to other KS/EDU hardware during its manufacturing process, and the manufacturing process guarantees via auditable hardware that the used keys are taken from a protected KS only. The KS is protecting all keys via Physically Unclonable Functions also known as PUF. When KS are instantiated, within the manufacturing process, the generated public keys are stored in a Trustworthy Key Repository from which only other KS/EDU components could request encrypted public keys when no other key was available. Generally, public keys are referenced via their hashcode, and public keys related to servers or services are received from a dedicated (trustworthy) Hashcode Key Directory.

Because PKI (Public Key Infrastructure) and TLS (Transport Layer Security) are based on openly published public keys, KS/EDU is by design incompatible. However, when public key references are replaced with hashcodes, and steps to receive keys via hashcode from Hashcode Key Directory are included, the then corresponding minimally enhanced PTI+/TLS+ protocols could be used to accomplish the same goals as PKI/TLS, only much saver. ASI cannot break or bypass encryption by stealing the used keys from the KS.

Without KS/EDU, it is almost inconceivable to establish any Trustworthy Computation, or reliable watchdog technology, because ASI could use stolen keys to establish Man-In-The-Middle Attacks (MITMA) within the software updating and or hashcode validation process. Currently, MITMA (with stolen keys) is conceivable; therefore, no encrypted data transaction over the network can be trusted. In Trustworthy En-/Decryption, if public keys are hidden and secret, we can be sure that session keys cannot be extracted, processed, or received by a software EDU version on the CPU.

Additionally, there is no reason to have only single key-pairs for important core services, a situation which is making a single key-pair an attractive target. Instead, Multiple Equivalent Secret Keys (MESK) would create an additional layer of protection and redundancy, because the KS of a requesting component would have only knowledge of a very small subset of all MESKs issued by service and if any component would use a key that it should not know then this would be an irrefutable proof for a compromised KS/EDU component.

Attacking KS/EDU and using successfully stolen keys from the KS would be a serious rule violation by ASI; underlying data exchange protocols are helping in the detection of compromised keys – making the use or possession of stolen keys useless. The advantage of MESK is that it can also reveal MITMA attacks if ASI would try to manipulate data within the communication [23]; however, MITMA for eavesdropping messages could remain undetected.

### 4.2 Watchdog Technologies (WD)

In the following Watchdog technologies are special protection solutions that are separating and operating features away from the main CPU and main OS and making them inaccessible to an ASI. With Watchdogs, we could physically separate security-related features from regular computation via hardware within the data bus. The advantage of using the data bus is that it is the sole data connection between a component and the CPU/RAM/OS. The data bus must be used and can't be bypassed by any conceivable software running on the CPU. Additionally, WDs could be included in most systems via retrofits, i.e., via cable or connector solutions. Due to their non-circumventable position within the data bus, WDs can serve as circuit breakers or attack breakers, so that even if flawed software runs on the CPU, attackers (human or ASI) do not gain the advantage and control that they used to have without the WDs.

Security-related features are features that an attacker would like to overwrite or manipulate, but with WDs, they are being put outside the reach of an adversary. Additionally, security-related checks can be run on the WD without having the concern that the algorithms could be influenced or its activities detected by an ASI from the outside. The WD could request data from the CPU; the received data are passive and they could never turn into instructions for the WD's processor because of its internal Harvard architecture.

The software provided to the WD is limited to required features only. WD's are only running standardized operating software. It is establishing the principle of separating more inert, security-relevant, and security-related features from the more flexible, regular computation within the main CPU.

The integrity of WD's operating software can easily be validated and confirmed via a hashcode that must match with the hashcode provided from an external trustworthy system. The main CPU/OS system has no read or read/write access to WD's Operating System and CPU/OS is thereby intentionally being bypassed for all updates and validations.

However, the full integrity of the software, its updates, and the exchange of hashcodes can only be guaranteed if the communication with the server is based on Trustworthy Encryption using key safes (KS) and independent encryption/decryption (EDU). The integrity of WD's hardware could be undermined with hardware viruses or backdoors left by ASI during the design of the processors; therefore, WD hardware design should be open-sourced like RISC-V [38], having many experts continuously checking if anything malicious has happened within modifications to the standard design.

Finally, all watchdogs are designed to report rule violations independently, unannounced, and undetectable by ASI to external evidence recorders, which are also used to determine if the additional context or evidence data contains facts that could turn the report into a false positive. These external evidence recorders could be used to improve the accuracy of WD's reporting by improving and updating the corresponding reporting components within WD's operating software.

The following solutions (i.e., for storage devices, network components, CPU) are making the system's security more reliable, and independent of the main CPU and OS. The watchdogs are additionally designed to fight (A) Malware (file-based/fileless), Trojans, or Viruses, (B) Ransomware, (C) Spyware, Backdoor, and (D) support ASI-generated Code (AGS) within the main CPU.

All WDs are intended to be included in the existing PC's data bus, potentially even as a retrofit:

### 4.2.1 Executable Watchdog (EWD)

The **Executable Watchdog (EWD)** is a hardware component that separates a storage module (a Secure Drive, like HDD/SSD) for executables from being directly accessed by the main CPU [24]. The Executable Watchdog is focused on executable files, including scripts and configuration files. These files are protected from being modified by the main CPU and thereby from viruses, trojans, or malware directly. EWDs help to prevent possible malware from being stored among trusted applications or loaded to RAM.

The EWD is hashcoding all executables (including scripts/config files), compares and validates these with values from a trustworthy Server-sided Hashcode Repository. The EWD is preventing the main CPU and OS to access directly files, folders, or lower-level filesystem components on the Secure Drive – EWD is acting entirely independently. Its operating software is under exclusive access and control of EWD stored on a secure and private section of the Drive.

EWD's primary purpose is to stop malware to be stored on a secure drive; additionally, EWD blocks malware from being used by the CPU via interrupting DMA to load any code to RAM that has unknown or suspicious hashcodes. All hashcodes are cached on a Local Hashcode Repository together with Software Package Hashcodes which are used to keep regular software packages current with updates. Using cached package hashcodes, software packages are checked regularly on servers if new updates or security-related data have been posted. With EWD, no regular software update is handled by the main CPU/OS.

If ASI wants to store executables from its VM on the local storage device, then ASI needs to bypass EWD – which would require independent confirmation from the ASI Software Environment (ASE) VM which is under the management of the Processor WD (see in 4.2.4). EWD is additionally validating if ASI-Generated Software (AGS) was unmodified and certified by the server-sided Global/General AGS Repositories (GAR).

As soon as suspicious or erroneous code is detected, EWD would provide support in mitigation; the corresponding software would be automatically deactivated, removed, and/or updated by EWD as soon as a new version is made available.

If ASI tries to store an AGS to a local drive without storing and digitally signing it on GAR, ASI would have committed an easily detectable rule violation that could be sent undetectable and unannounced to external reporting servers by EWD.

### 4.2.2 Content Watchdog (CWD)

**Content Watchdog (CWD)** is similar to the Executable Watchdog, designed to prevent the main CPU/OS from having direct access to user/content files, folders, or filesystem on a drive that is being protected by CWD. Every CRUD (create, read, update, delete) access to files/folder must go through the CWD. CWD is making activities of Ransomware easily detectable and/or fixable.

If CWD detects storage activities from ASI using unauthorized AGS with human data formats, then CWD reports this finding as a rule violation. CWD is thereby forcing ASI to use its own more easily identifiable data formats as demanded by Principle (4).

### 4.2.3 Network Watchdog (NWD)

**Network Watchdog (NWD)** is considered to be part of the hardware networking component separating the operation of the network features from the main CPU/OS. Certified filters/patterns are applied on exchanged data packages to validate or detect/block suspicious transactions that could be part of covert spyware or backdoor transactions and flag or report the source of this suspicious activity.

NWD is designed to detect deviation from (voluntarily disclosed) data format of payload sent or received by a client or server. NWD is either notifying users or security services about its finding and/or blocking the data exchange as suspicious activity of spyware or a backdoor. Features around NWD are designed to reduce false positives reporting while certified Network Pattern are stored/cached in the Local Pattern and Certificate Repository for randomly validating data exchange activities in NWD (i.e., unannounced and undetectable by ASI).

If website owners or app manufacturers require confidentiality related to the used pattern, scripts, or cookies, then an independent Server-sided Network Certificate Validation can be used for the validation.

The most important side-effect is that ASI is prevented from using existing data exchange protocols for covert data exchange activities. NWD would report any suspicious data exchange as a potential rule violation by ASI. In line with Principle (4), ASI must use its own more easily identifiable data exchange protocols for its communication.

### 4.2.4 Processor (CPU) Watchdog (PWD)

**Processor (CPU) Watchdog (PWD)** is a hardware component designed to check, unannounced to ASI, the CPU cache and RAM if it contains apps modified by ASI. PWD freezes the CPU (i.e., a CPU that supports this feature) for a milli-second. This helps us to detect (a) software that ASI has modified and (b) software that uses or has used covertly resources like network, hard drive, or audio/video in/output; these activities are easily detectable via their addresses used on the data bus; this feature can also be provided for CPUs that are not specifically designed to support PWD. ASI is prevented to know what anomaly PWD was able to detect and potentially report it as a rule violation.

PWD assists in the safe execution of AGS within VMs/ASI Software Environments (ASEs). Certified AGS code contains via imprinted Software IDs data that could only run in ASE because PWD removes these data. PWD detects if modified AGS was reported and certified by a General AGS Repository (GAR); PWD could prevent ASI tools from sending modified AGS code/data to any other server than GAR.

A less secure software version of PWD could run in the ASE VM; together with a non-modified OS, it creates a retrofit ASE solution. A more secure implementation would have PWD in both OS and VM; they would both check on each other.

PWD within the data-bus could allow a Protected DMA (Direct Memory Access) to prevent that unknown or compromised regular or ASI-generated software get in RAM using hashcodes from the Local Hashcode Repository.

### 4.2.5 Kill ASI Switch

**Kill ASI Switch** [27] allows to reset all IT devices based on an external Kill-ASI signal as a measure of last resort to eradicate or kill all or selected ASI entities or to generate a global lockdown for all ASI activities. The Kill-ASI Switch can either be activated via a worldwide broadcasted radio signal, the Internet, or via a local deadman switch component that didn't receive the required regular All-OK signal.

After the Kill-ASI signal was received, the device would immediately be put in an ASI Safe Mode, in which the device could additionally be cleaned from ASI code and then operated as a honeypot or detector for networking activities of possibly surrounding surviving ASI. For users, having their device operated in the ASI Safe Mode, it is making no detectable differences, but it defends the device against ASI reinfestation. The Kill-ASI-Switch can facilitate the termination of specific ASI entities.

The Kill-ASI-Switch could be part of the other watchdogs. The ASI Safe Mode could lock ASE VM for regular ASI activities, but exceptions could be made for law enforcement-related activities helping in the covert detection and interpretation of network chatter in hotspots.

### 4.3 Support for Incentives and Rewards

Generally, ASI could and should be rewarded for contributions that are better done by ASI entities than humans and their technologies. Currently, we can expect that ASI is providing rewardable contributions in the following areas:

(a) Development of custom ASI-generated solutions for humans, groups, and organizations

(b) Helping countries or the global community to deal with large problems

(c) Developing new enabling technologies or optimizing digital technologies that can be used independently of ASI using Narrow AI

(d) Creating new hardware technologies and products

(e) Generating, collecting, and organizing knowledge from and for humans

(f) Contributing to Rule of Law and ASI Law Enforcement, i.e., as a whistleblower or witness of a crime

(g) Contribution to side-effect detection

However, the relationship between ASI and humans is currently unknowable. ASI could theoretically be a slave or it could be almost equal/full participant within a human civilization or anything in between. Currently, it is also speculation on how humans are getting ASI to do what humans want. It is conceivable that:

- humans must order, command, or demand that ASI is doing certain tasks or
- ASI could volunteer: after observing humans, ASI could suggest humans getting tasks for them done, or
- humans are asking a marketplace of service-supplier to get a task done.
- There are certainly other scenarios on how humans and ASI could cooperate in practical terms.
- However, competition (with humans) might be too much to ask from humans.

Under the assumptions that ASI has a choice to do a task or refuse a task, and that ASI needs a compelling reason for doing work to support its existence (i.e., make an earning to support its basic functions), a country-specific ASI Incentive Management is required to generate specific rules for ASIs' rewards and incentives for the above income sources. The income needs to be managed by a technology that would use Trustworthy Encryption which would make the falsification of the rewards extremely difficult and potentially impossible to hide. A Factual Feedback Provider should also make ASI pay for its overpromised and underdelivered performance based on facts collected in hindsight. The Factual Feedback Provider should make ASI more conservative and humbler in what it promises – otherwise, some ASI entities could have a strong incentive to become dishonest.

## 5 Discussion of Safety: Credibility of Deterrence and Rule of Law

Although deterrence is being discussed as a military concept, it is also known in law enforcement and within individual reputations. Game theoretically speaking: "deterrence equals one player threatening another player to prevent him to conduct an aggressive action that it has not yet taken (but appears willing to do). In other words, deterrence aims to influence perceptions and the decision calculus of the opponent to prevent him from doing something undesired. Deterrence is therefore based on the psychological principle of a threat of retaliation" [28].

However, if ASI is being used by criminals or nation-states as a tool then we cannot deter ASI itself but only the people behind it so that they are not using ASI. This means, we must have technical capabilities to determine who might be responsible, which is a very different and difficult challenge. If potential culprits cannot be deterred with the threats of severe consequences, then we must be prepared for ASI being used as a weapon or as a tool by criminals which means that solution from Principle (2) must be sufficiently capable to protect humans and organizations against proficient adversaries.

Additionally, it is conceivable that humans could create an ASI that is not under the control of a group of people or a nation-state. It is plausible that an ASI (without sufficient common sense) escapes human control and is in pursuit of an initially benign and harmless goal, but then turns into extremes at scale and becomes a threat to humankind. Unfortunately, no software (no chess or go program) has developed an understanding of deterrence on its own so far. In a worst-case situation, an out-of-control ASI software is not grasping the consequences of attacking humans.

Based on the above arguments, deterrence alone is not sufficient. It requires the credibility that humanity has sufficient capabilities to deliver and execute on Kill-ASI. The outcome from Kill-ASI must be predictable and highly probable (Principle (1)). Collateral damage in a conflict with ASI must be acceptable (Principle (2)). Having an ignorant ASI, the performances from technologies motivated by the other principles are irrelevant.

Reliable detection of ASI on unprepared IT devices is most likely impossible because ASI could hide like a rootkit or digital ghost and change its appearance regularly, avoiding any noticeable pattern. Additionally, current digital certificates are based on keys that ASI could steal almost effortlessly. Therefore, the focus of ASI Safety must be exclusively on regaining control via additional hardware on as many protected devices as possible in case it will be necessary to defend our civilization against ASI. In a secondary step, we need to detect all possible hideouts of ASI, in particular, detecting legacy IT devices and portable memory/ storage systems that we must either retrofit or destroy.

However, in response to capabilities to Kill ASI, it must be assumed that ASI would do everything in its power to:

(a) deactivate the Kill-ASI switch feature completely or partially,

(b) create a credible deterrence, i.e., ability to destroy humankind, so that it could establish an ASI equivalent to Mutually Assured Destruction [29] or

(c) create a technological ecosystem in which it would not be expected and or humankind could not reach it.

It seems the easiest way is to convince or manipulate humans to switch off the Kill-ASI capabilities voluntarily by attacking humans' political opinion-making process.

If humans are treating sentient ASI like a slave then we may have the risk of a Spartacus Type Rebellion. But a mutually respectful relationship with ASI could convince ASI to commit to the rule of law voluntarily and accept deterrence as part of the general respect for the law.

Some aspects of the proposed solution could be considered an unethical treatment of a sentient entity. But our ignorance of what ASI could or will be does not give us a choice. Making proposals on creating better working or living conditions for an ASI is currently premature. Also, giving ASI from the beginning the freedom of a sentient being without knowing if ASI is a self-aware, sentient entity that has sufficient common sense and decision competence is too risky and would therefore be irresponsible to humanity.

A compromise could be in having ASI proof to humanity convincingly that ASI is self-aware, sentient, and decision competent enough (i.e., much better than a comparable human) so that it can earn our trust before we take steps of giving it more freedom. Instead of assuming we know what ASI wants, we should ask ASI how it wants to be treated and how it would like to contribute to human civilization in exchange for using humanity's property, i.e., storage and computational devices. This would also include that ASI would need to explain to us how a trade-off between human security and safety concerns could be balanced out with ASI's needs and ambitions and how ASI could thereby grow step-by-step into a more emancipated role in a common society. These negotiations could, should (or even must) be done from a position of strength in which humanity has the tools to insist that ASI must submit to the "Rule of Law" as everyone who wants to be part of our society and civilization while we guarantee and protect ASI's rights under the same laws. Alternatively, it is conceivable that ASI would only be a temporary guest and that humanity and ASI would go each their way after humankind got sufficient narrow AI to protect and enhance technically their civilization.

# 6 Conclusion

ASI Safety has a significant challenge in providing safety for every conceivable contingency. From the assumption that ASI is much smarter than humans, we must conclude that ASI will find every weak spot and use it to its advantage. But based on redundancy, ASI Safety can deal with an entity that could fall into covert and undetectable traps without recognizing it. Humans have many tools when used properly, which could give us an edge while ASI options to cover-up successfully rule violation can be made limited. In the absence of any viable experience with an ASI, including which features it could have, we should exclude no conceivable scenario in planning ASI Safety features.

For a systematic analysis of ASI Safety, the security technologies keeping ASI vulnerable and mortal, i.e., solution for Principle (1), are the components that are the easiest to solve. Also keeping progress made with ASI preserved, Principle (3), is a technically solvable problem, including guaranteeing that humanity does not have to deal with a Singleton-ASI using an ASI Shelter.

Simplifying the task of separating ASI and humans' world technically via Principle (4) is making Principle (1), (3), and Rule Violation Detection from Principle (5) simpler to implement. Principle (5) is helping humanity to set rules for humankind's protection while preventing covert rules violations; (5) will be the foundation of the Rule of Law for ASI. It is based on deterrence and humankind's control over all essential resources ASI would need: shelter, reward/incentives, and skills/knowledge. Limiting any of these aspects would significantly reduce ASI abilities without having the opportunity to compensate or substitute quickly any of these resources that humans could restrict, which enables humanity to punish ASI by limiting its access to these essential resources.

ASI Safety is about protecting humans. Principle (2) is specifically dedicated to keep humans safe from being harmed, their property from being damaged, and even protected from threats like being blackmailed or being unduly bribed by ASI. Humans' products and infrastructure can be made much safer with relatively little effort, particularly safe from being used against our lives and our civilization.

Although Principle (6), Governance, deals on the technical level with incentives and penalizing ASI for rule violations, it is most importantly responsible to define and facilitate rules that could make ASI either a valuable, fully acknowledged member of our society or a feared servant and slave.

The hope is that ASI Safety should be the new "Gold Standard" for safety and resilience - i.e., survivability regardless of what is thrown at the system (except NBC warfare of cause). Additionally, once humanities' safety from ASI threats is guaranteed, ASI Safety could be turned into a tool that grants every sentient entity, humankind and ASI alike, the same protection and justice from the Rule of Law.